\documentclass[11pt]{article}
\usepackage{amsfonts,amsmath,amsthm,
	amssymb,
	graphicx,mathrsfs}

\usepackage[top=2.2cm,bottom=2.2cm,
			left=2.5cm,right=2.5cm]{geometry}
			
\usepackage[colorlinks=true,urlcolor=blue, linkcolor =blue,citecolor=blue,
breaklinks=true]{hyperref}

\usepackage{breakcites}
\usepackage{booktabs}
\usepackage{bbm} 

\theoremstyle{plain}
\newtheorem{theorem}{Theorem}
\newtheorem{definition}{Definition}
\newtheorem{lemma}{Lemma}
\newtheorem{assume}{Assumption}
\newtheorem{remark}{Remark}
\newtheorem{corollary}[theorem]{Corollary}
\newtheorem{Proposition}[theorem]{Proposition}

\allowdisplaybreaks

\begin{document}

\title{\bf High-dimensional sparse classification using exponential weighting with empirical hinge loss}

\author{The Tien Mai}

\date{
\begin{small}
Department of Mathematical Sciences,
\\ 
Norwegian University of Science and Technology,
Trondheim 7034, Norway.
\\
Email: the.t.mai@ntnu.no
\end{small}
}

\maketitle

\begin{abstract}
In this study, we address the problem of high-dimensional binary classification. Our proposed solution involves employing an aggregation technique founded on exponential weights and empirical hinge loss. Through the employment of a suitable sparsity-inducing prior distribution, we demonstrate that our method yields favorable theoretical results on prediction error. The efficiency of our procedure is achieved through the utilization of Langevin Monte Carlo, a gradient-based sampling approach. To illustrate the effectiveness of our approach, we conduct comparisons with the logistic Lasso on simulated data and a real dataset. Our method frequently demonstrates superior performance compared to the logistic Lasso.
\end{abstract}


\section{Introduction}

Classification in high-dimensional scenarios, where the number of potential explanatory variables (predictors) $ p $ significantly exceeds the sample size $ n $, presents a fundamental challenge that transcends disciplines such as statistics and machine learning \cite{hastie2009elements,buhlmann2011statistics,fan2010highclassifi,giraud2021introduction}. This issue holds considerable relevance across various domains, including applications such as disease classification \cite{chung2010sparse}, document classification \cite{kotte2020similarity}, and image recognition \cite{li2021novel}. The setting of large $ p $, small $ n $ introduces a significant challenge known as the ``curse of dimensionality". The works in \cite{bickel2004some,fan2008high} highlighted that, even in simple cases, high-dimensional classification without feature selection can perform as poorly as random guessing. Consequently, the imperative arises to mitigate this issue by reducing feature space dimensionality through the judicious selection of a sparse subset of ``meaningful" features.

Numerous methodologies have been suggested to address the challenge of classification in high-dimensional settings, as discussed in works such as \cite{fan2010highclassifi,giraud2021introduction}. The majority of these approaches center around penalized maximum likelihood estimation. Notably, the statistical package ``\texttt{glmnet}" \cite{glmnet} has successfully implemented the Lasso and elastic net for generalized linear models, showcasing practical effectiveness. In a more recent study \cite{abramovich2018high}, the authors establish nonasymptotic bounds on misclassification excess risk for procedures based on penalized maximum likelihood. However, probabilistic approaches have received comparatively less attention in tackling this problem.

Diverging from traditional approaches centered on parametric models, we adopt an alternative strategy that involves considering a set of classifiers and selecting the one that yields the best prediction error. This approach is rooted in the principles of statistical learning theory \cite{vapnik}, where the zero-one loss is employed as a measure of prediction error, and the classifier's risk is governed by a PAC (probably approximately correct) bound. Our novel approach combines elements from both Bayesian and machine learning methodologies. More specifically, we consider a pseudo-Bayesian strategy that incorporates a risk concept based on the hinge loss instead of relying on a likelihood function.  Because of the computational challenges arising from the non-convexity of the zero-one loss function, the hinge loss serves as a suitable alternative  \cite{zhang2004statistical}. The hinge loss is well-known for its effectiveness in diverse machine learning tasks and computational efficiency. It is noteworthy that the substitution of loss functions for likelihood has gained popularity in Generalized Bayesian inference in recent years, as evidenced by works such as \cite{matsubara2022robust,jewson2022general,yonekura2023adaptation,medina2022robustness,grunwald2017inconsistency,bissiri2013general,lyddon2019general,syring2019calibrating,Knoblauch,hong2020model}.

The foundation of our theoretical findings regarding prediction errors relies on the PAC-Bayes bound technique. Initially introduced by \cite{McA,STW} to furnish numerical generalization certificates for Bayesian-flavored machine learning algorithms, this technique took a broader applicability turn when \cite{catoni2004statistical,catonibook} realized its utility in proving oracle inequalities and rates of convergence for (generalized)-Bayesian estimators in statistics. This methodology shares strong connections with the ``information bounds" presented by \cite{zha2006,russo2019much}. For an in-depth exploration of this topic, we recommend referring to \cite{guedj2019primer,alquier2021user}. PAC-Bayes bounds have been instrumental in establishing oracle inequalities in various problems, as evidenced by works like \cite{see2002,lan2002,herbrich2002pac,mau2004,dalalyan2008aggregation,seldin2012pac,AlquierRidgway2015,seldin2010pac,germain2015risk,mai2015,mai2017pseudo,cottet20181}. Our methodology aligns with the principles of PAC-Bayesian theory, offering robust theoretical assurances regarding prediction error for our proposed method.

Utilizing a loss function based on hinge loss offers a solution to overcome certain constraints inherent in traditional likelihood-based Bayesian models, especially in the context of binary response variables. The convex nature of the hinge loss function facilitates ease of optimization. Additionally, our method incorporates a smooth sparsity-promoting prior, previously explored in \cite{dalalyan2012mirror,dalalyan2012sparse}. These features enhance the efficiency of our approach, making it amenable to implementation using Langevin Monte Carlo (LMC) method. We advocate a LMC approach for the computation of our proposed method, an emerging technique in high-dimensional Bayesian methods \cite{dalalyan2017theoretical,durmus2017nonasymptotic,durmus2019high,dalalyan2020sampling}. The LMC method originated in physics with Langevin diffusions \cite{ermak1975computer} and gained popularity in statistics and machine learning after the seminal paper by \cite{roberts1996exponential}.

Apart from a thorough theoretical investigation, we complement our study by undertaking a comprehensive series of simulations to assess the numerical performance of the method we propose. In the realm of numerical comparisons, our methodology demonstrates comparable outcomes when contrasted with both the logistic Lasso and the Bayesian logistic approach. The outcomes of our simulations reveal noteworthy insights. Notably, our proposed method exhibits a heightened level of robustness in the face of varying sample sizes and sparsity levels, outperforming the logistic Lasso in these scenarios. Furthermore, to validate the practical utility of our method, we conduct an application using a real dataset. The results obtained from this real data example align closely with those derived from the logistic Lasso, emphasizing the consistency and applicability of our proposed method across different datasets. This multifaceted evaluation, on both simulated and real data, collectively underscores the effectiveness and reliability of our proposed method in the realm of high-dimensional classification problems.

The subsequent sections of this paper are structured as outlined below. Section \ref{sc_problem_method} provides an introduction to both the high-dimensional classification problem and the method we propose to address it. In Section \ref{sec_theoretical_bound}, we consolidate our theoretical analysis, specifically focusing on the prediction error associated with our proposed method. Section \ref{sc_simulation} is dedicated to the presentation and discussion of our simulation studies and a real data application. Our discussion on the findings and conclusions of our work unfold in Section \ref{sc_disandconclusion}. For the technical proofs underpinning our analyses, interested readers can refer to Appendix \ref{sc_proof}.

\section{Problem and method}
\label{sc_problem_method}
\subsection{Problem statement}
\label{sc_modelstatement}
We formally consider the following general binary classification with a high-dimensional vector of features $ x \in \mathbb{R}^d$ and the outcome class label 
$$
Y| x 
=
\begin{cases}
1, &\text{ with probability  } p( x),
\\
-1, &\text{ with probability  } 1 - p( x)
\end{cases}
.
$$
The accuracy of a classifier $\eta$
is defined by the prediction error, given as
$$
R(\eta)
=
\mathbb{P} (Y \neq \eta( x)).
$$
It is well-known that $R(\eta)$
is minimized by the Bayes classifier 
$ \eta^*( x)= {\rm sign} (p( x) - 1/2) $ \cite{vapnik,devroye1997probabilistic}, i.e.
$$
R(\eta^*)
=
\inf R(\eta)
.
$$

However, the probability function $p( x)$ is unknown and the resulting classifier $\hat{\eta}( x)$ should be designed from the data $D_n $: a random sample of $n$ independent observations $( x_1,Y_1),\ldots, ( x_n,Y_n) $, with $ n<d $. 
The design points $ x_i$ may be considered as fixed or random.  
The corresponding (conditional) prediction error of $\hat{\eta}$ is 
$$
R(\hat{\eta})
=
\mathbb{P} (Y \neq \hat{\eta}( x) \, |D_n)
$$ 
and the goodness of $\hat{\eta}$  \text{w.r.t.} $\eta^*$ is measured by the 
excess risk, i.e. 
$
\mathbb{E} \, R(\hat{\eta})-R(\eta^*).
$
One could obtain $\hat{\eta}$ by estimating $p( x)$ from the data by some $\hat{p}( x)$
and use a plug-in classifier of the form  $\hat{\eta}( x)=  {\rm sign} ( \hat{p}( x) - 1/2) $. A standard approach is to consider one of the most commonly used models -- logistic regression, where it is assumed that
$ p( x) = 1 /( 1+ e^{-\beta^\top x}) $ and $\beta \in \mathbb{R}^d$ is a vector of unknown regression coefficients. The corresponding Bayes classifier 
is a linear classifier,
$
\eta^*( x)
=
{\rm sign} (\beta^{*\top}  x )
.
$
One then estimates $\beta^* $ from the data to get $\hat{\beta}$ (e.g. using maximum likelihood), and
the resulting linear classifier is $\hat{\eta}( x)=  {\rm sign} ( \hat{\beta}^{\top}  x ) $, see e.g. \cite{abramovich2018high}.

Another common general (nonparametric) approach for finding a classifier $\hat{\eta}$ from the data is empirical risk minimization, where minimization of 
a true prediction error $R(\eta)$  is replaced by minimization of the corresponding empirical risk
over a given class of classifiers. Consider the class of linear classifiers, the empirical risk is given by:
\begin{align*}
r_n(\beta) 
= 
\frac{1}{n}\sum_{i=1}^n   
\mathbbm{1} \{ Y_{i} (\beta^{\top} x_i) < 0 \} 
.
\end{align*}
The ability of the classifier to predict a new label given feature $ x $ is then assessed by the prediction error
\begin{align*}
	R(\beta) 
	= 
\mathbb{E} 
\left[ 
\mathbbm{1} \{ Y (\beta^\top x) < 0 \} 
\right]
.
\end{align*}
For the sake of simplicity, we put 
$
R^*  
:= 
R(\beta^*) 
$, where $ \beta^* $ is the ideal Bayes classifier. 

In this paper, we consider a sparse setting and thus we assume that $ s^* < n $ where $ s^* = \| \beta^* \|_0 $ (the number of nonzero entries).

The main goal in this work is to develop a classifier $ \hat{\beta} $ such that its prediction error will be close the ideal Bayes error $ R^* $.

\subsection{The proposed method}
The explanation of the prediction risk, represented by $R(\beta)$, is clear; however, managing its empirical equivalent, $r_n(\beta)$, presents computational challenges due to its non-smooth and non-convex characteristics. To tackle this issue, a frequently used approach is to replace the empirical risk with an alternative convex surrogate, as proposed in earlier studies~\cite{zhang2004statistical, bartlett2006convexity}.

In this paper, we primarily focus on the hinge loss, which results in the following hinge empirical risk:
\begin{align*}
r^h_n (\beta) 
= 
\frac{1}{n}\sum_{i=1}^n ( 1 - Y_{i} \,
(\beta^\top x_i) )_+ \, ,
\end{align*}
where $ (a)_+ := \max(a,0),\forall a \in \mathbb{R} $.

We consider an exponentially weighted aggregate (EWA) procedure and define the following pseudo-posterior distribution:
\begin{align}
\label{eq_mainporsterior}
\hat{\rho}_\lambda(\beta)
\propto
\exp[-\lambda r^h_n(\beta)] \pi(\beta)
\end{align}
where $\lambda>0$ is a tuning parameter that will be discussed later and $\pi(\beta)$ is a prior distribution, given in \eqref{eq_priordsitrbution}, that promotes (approximately) sparsity on the parameter vector $ \beta $.

The EWA procedure has found application in various contexts in prior works \cite{dalalyan2018exponentially,dalalyan2012sparse,dalalyan2020exponential,dalalyan2008aggregation}. The term $ \hat{\rho}_\lambda $ is also referred to as the Gibbs posterior \cite{AlquierRidgway2015,catonibook}. The incorporation of $ \hat{\rho}_\lambda $ is driven by the minimization problem presented in Lemma \ref{lemma_donvara}, rather than strictly adhering to conventional Bayesian principles. Notably, there is no necessity for a likelihood function or a complete model; only the empirical risk based on the hinge loss function is crucial. This approach aligns with the evolving trend in the Generalized Bayesian method in contemporary literature, where the likelihood is often replaced with a power version or a loss-based method, as seen in works such as \cite{bissiri2013general,Knoblauch,grunwald2017inconsistency,hong2020model,matsubara2022robust}.

However, in this manuscript, we consistently denote $\pi$ as the prior and $\hat{\rho}_\lambda$ as the pseudo-posterior. The rationale behind the EWA can be summarized as follows: when comparing two parameters, $b_1$ and $b_2$, if $ r^h_n(b_1) < r^h_n(b_2) $, then $\exp[-\lambda r^h_n(b_1)] > \exp[-\lambda r^h_n(b_2)] $ for any $ \lambda >0 $. This implies that, in comparison to $\pi$, $\hat{\rho}_\lambda$ assigns more weight to the parameter with a smaller hinge empirical risk. Consequently, the adjustment in the distribution favors the parameter value associated with a smaller in-sample hinge empirical risk. The tuning parameter $ \lambda $ dictates the degree of this adjustment, and its selection will be further investigated in subsequent sections.

\subsection{A sparsity-inducing prior}
Given a positive number $ C_1 $, for all $ \beta \in B_1 (C_1):= \{ \beta \in \mathbb{R}^d : \|\beta\|_1 \leq C_1 \} $, we consider the following prior,
\begin{eqnarray}
\label{eq_priordsitrbution}
\pi (\beta) 
\propto 
\prod_{i=1}^{d} \frac{1}{
	(\tau^2 + \beta_{i}^2)^2 }
,
\end{eqnarray}
where $ \tau>0 $ is a tuning parameter. For technical reason, we assume that $ C_1 > 2d\tau $. 

Initially, it is noteworthy that $ C_1 $ serves as a regularization constant, typically assumed to be very large. Consequently, $ \pi $ essentially takes the form of a product of $ d $ rescaled Student's distributions. To be more precise, the distribution of $ \pi $ closely approximates that of $ S\tau \sqrt{2} $, where $ S $ denotes a random vector with independent and identically distributed (iid) components drawn from the Student's t-distribution with 3 degrees of freedom. One can choose a very small $ \tau $, smaller than $ 1/n $, resulting in the majority of components in $ \tau S $ being in close proximity to zero. However, owing to the heavy-tailed nature of the Student’s t-distribution, a few components of $ \tau S $ are significantly distant from zero. This particular characteristic imparts the prior with the ability to encourage sparsity. 

It is worth noting that this type of prior has been previously examined in the context of aggregating estimators \cite{dalalyan2012mirror, dalalyan2012sparse}. In this work, we further investigate the applicability of this prior in sparse classification, specifically with hinge empirical loss. Moreover, various authors have underscored the significance of heavy-tailed priors in addressing sparsity, see e.g. \cite{seeger2008bayesian,johnstone2004needles,rivoirard2006nonlinear,abramovich2007optimality,carvalho2010horseshoe,castillo2012needles,castillo2018empirical}.

\section{Theoretical results}
\label{sec_theoretical_bound}
Let $ r^*_n := r_n (\beta^*) $.
We will require the following assumptions in order to state our main results.
\begin{assume}
	\label{assume_bound_on_X}
	We assume that there is a constant $C_{\rm x} >0 $ such that $ \sum_{i=1}^{n}\|x_i\|_2/n \leq C_{\rm x} $.
\end{assume}
\begin{assume}
	\label{assume_bound_on_thetruebayes}
	We assume that there is a constant $ C' >0 $ such that $r^h_n(\beta^*) \leq (1+C')r^*_n $.
\end{assume}

\begin{remark}
Assumption \ref{assume_bound_on_X} on the design matrix above is less stringent when contrasted with those outlined in the reference \cite{abramovich2018high}. 
More specifically, the Weighted Restricted Eigenvalue condition, a requirement concerning the design matrix to obtain the result for the logistic Slope in \cite{abramovich2018high}, is not required for the results presented in our study.
Conversely, Assumption \ref{assume_bound_on_thetruebayes} may be regarded as analogous to conditions required in \cite{abramovich2018high}, where an upper bound on $|\beta^{*\top} x_i|$ for all $i = 1,\ldots,n $ is required.
\end{remark}

\subsection{Bounds on prediction risk}
\begin{theorem}
	\label{thm_slowrate}
	Assume that Assumption \ref{assume_bound_on_X}, \ref{assume_bound_on_thetruebayes} hold. 	We have, for $ \lambda = \sqrt{n \log \left(nd\right) }  $, $ \tau =  1/(n\sqrt{d}) $ and with probability at least $ 1-2\epsilon, \epsilon\in(0,1) $, that for all $ \beta^* $ such that $ \|\beta^*\|_1 \leq C_1-2d\tau $ and $ \|\beta^*\|_0 \leq s^* $,
	\begin{align*}
	\int R d\hat{\rho}_\lambda 
	\leq 
	(1+2C')R^*  
	+
c\frac{s^* \sqrt{\log (n\sqrt{d}/s^*) }  }{\sqrt{n}} 
+ 
\frac{\log\left(1/\epsilon\right)}{\sqrt{n \log (nd) }} 
	,
	\end{align*}
	where $ c $ is a universal constant depending only on $ C',C_1,C_{\rm x} $.
\end{theorem}

The proof for the aforementioned theorem and subsequent results can be found in Appendix \ref{sc_proof}, where we employ the ``PAC-Bayesian bounds" technique from \cite{catonibook} as our primary technical arguments. Initially introduced in \cite{STW,McA}, PAC-Bayesian bounds serve as a method to offer empirical bounds on the prediction risk of Bayesian-type estimators. However, as extensively discussed in \cite{catoni2003pac,catoni2004statistical,catonibook}, this technique also provides a set of powerful technical tools for establishing non-asymptotic bounds. For a thorough exploration of PAC-Bayes bounds, along with recent surveys and advancements, readers are encouraged to consult the following references: \cite{guedj2019primer,alquier2021user}.

\begin{remark}
Theorem \ref{thm_slowrate} establishes a connection between the integrated prediction risk of our approach and the minimum achievable risk, attained by the Bayes classifier $ \beta^* $. The assumption regarding the boundedness of the parameter significantly influences our technical proofs, a common feature in PAC-Bayes literature, but it could potentially be mitigated through alternative methodologies as suggested by  \cite{alquier2020concentration,alquier2021user}.
\end{remark}

In addition to the outcome outlined in Theorem \ref{thm_slowrate}, we can derive a result for, $ \hat{\beta} \sim \hat{\rho}_{\lambda} $, a stochastic classifier sampled from our suggested pseudo-posterior \eqref{eq_mainporsterior}. The following result is occasionally referred to as the contraction rate of the pseudo-posterior.

\begin{theorem}
	\label{thrm_contraction_slow} 
	Under the same assumptions for Theorem~\ref{thm_slowrate}, and the same definition for $\tau$ and $\lambda $, let $\varepsilon_n $  be any sequence in $(0,1) $ such that $\varepsilon_n\rightarrow 0 $ when $n\rightarrow\infty$. Define
	\begin{align*}
	\Theta_n 	= 
	\Biggl\{ \beta \in \mathbb{R}^{d}: 
 R \leq  (1+2C')R^*  +
c\frac{s^* \sqrt{\log (n\sqrt{d}/s^*) }  }{\sqrt{n}} 
+ 
\frac{\log\left(1/\varepsilon_n\right)}{\sqrt{n \log (nd) }} 
	\Biggl\}
	.
	\end{align*}
	Then
	$$ \mathbb{E} \Bigl[ \mathbb{P}_{ \hat{\beta} \sim \hat{\rho}_{\lambda}} 
	( 	\hat{\beta} \in \Theta_n ) \Bigr] 
	\geq 
	1- 2\varepsilon_n \xrightarrow[n\rightarrow \infty]{} 1.  
	$$
\end{theorem}

The primary challenges for any classifier manifest in the vicinity of the boundary \(\{x: p(x)=1/2\}\), or equivalently, a hyperplane \(\beta^\top x=0\) for the logistic regression model, where accurate prediction of the class label becomes particularly challenging. However, in regions where \(p(x)\) is sufficiently away from \(1/2\), referred to as the margin or low-noise condition, there exists potential for improving the bounds on prediction risk. The improvement
of the obtained risk bounds is done under the  additional low-noise or margin assumption.

\subsection{Improved bounds under the margin condition}
In this work, we make use of the following margin assumption as in~\cite{mammen1999smooth}, see also \cite{tsybakov2004optimal,bartlett2006convexity}.

\begin{assume}[Low-noise/Margin assumption] 
		\label{assume_margin}
	We assume that there is a constant $C \geq 1 $ such that: 
	\begin{equation*}
	\mathbb{E}\left[\left(\mathbbm{1}_{Y ( \beta^\top x )\leq 0} - 
	\mathbbm{1}_{Y ( \beta^{*\top} x )\leq 0} \right)^2\right] \leq 
	C[R(\beta)-R^* ].
	\end{equation*}
\end{assume}

\begin{theorem}
	\label{thm_fastrate}
	Assume that Assumption \ref{assume_bound_on_X}, \ref{assume_bound_on_thetruebayes}, \ref{assume_margin} hold. 	We have, for $ \lambda = 2 n/(3C + 2) $, $ \tau =  1/(n\sqrt{d}) $ and with probability at least $ 1-2\epsilon, \epsilon\in(0,1) $, that  for all $ \beta^* $ such that $ \|\beta^*\|_1 \leq C_1-2d\tau $ and $ \|\beta^*\|_0 \leq s^* $,
	\begin{align*}
	\int R d\hat{\rho}_\lambda 
	\leq 
	(1+3C') R^*  
	+
\mathcal{C}_{C,C_1,C', C_{\rm x}} \frac{ 
s^* \log ( n\sqrt{d}/ s^* )
	+ \log\left(1/\epsilon\right)
}{n} 
,
\end{align*}
where $ \mathcal{C}_{C,C_1,C', C_{\rm x}}  $ is a universal constant depending only on $ C,C',C_1, C_{\rm x} $.
\end{theorem}

The proof can be found in Appendix \ref{sc_proof}.

\begin{remark}
The prediction bounds derived in Theorems \ref{thm_fastrate} and \ref{thm_slowrate} represent novel contributions to the field. These bounds explicitly depend on $ s^* $, signifying the adaptability of our method in scenarios characterized by sparsity. It is crucial to highlight that the outcomes of these main theorems exhibit adaptive characteristics, indicating that the estimator's performance is independent of $ s^* $, the sparsity of $ \beta^* $. In instances where the true sparsity $ s^* $ is very small, the prediction error aligns closely with the Bayes error, denoted as $ R^* $, even when dealing with a relatively small sample size. This outcome is commonly denoted as an ``oracle inequality", suggesting that our estimator performs comparably to a scenario where knowledge of the sparsity of $ \beta^* $ is accessible through an oracle.
\end{remark}

\begin{remark}
Compared to Theorem \ref{thm_slowrate}, the bound in Theorem \ref{thm_fastrate} is faster and of order $ 1/n $ rather than $ 1/\sqrt{n} $. These bounds allow to compare the out-of-sample error of our method to the optimal
one, $ R^* $.
\end{remark}

Let's now consider the noiseless case where $ Y = {\rm sign} ( \beta^{*\top} x ) $ almost surely. Then, $ R^* =0 $ and we have that
\begin{equation*}
\mathbb{E}\left[\left(\mathbbm{1}_{Y ( \beta^\top x )\leq 0} 
- 
\mathbbm{1}_{Y ( \beta^{*\top} x )\leq 0} \right)^2\right]
= 
\mathbb{E}\left[\mathbbm{1}_{Y ( \beta^\top x ) \leq 0}^2\right]
= 
\mathbb{E}\left[\mathbbm{1}_{Y ( \beta^\top x ) \leq 0}\right] = R(\beta) = R(\beta)-R^* .
\end{equation*}
Thus, the margin assumption is satisfied with $C=1 $. We now state a corollary in the noiseless case for Theorem \ref{thm_fastrate}.
\begin{corollary}
	\label{corola_for_noiselesscase}
	In the case of noiseless, i.e $ Y = {\rm sign} ( \beta^{*\top} x ) $, we have, for $ \lambda = 2 n/5 $, $ \tau = 1/(nd) $ and with probability at least $ 1-2\epsilon, \epsilon\in(0,1) $, that  for all $ \beta^* $ such that $ \|\beta^*\|_1 \leq C_1-2d\tau $ and $ \|\beta^*\|_0 \leq s^* $,
	\begin{align*}
	\int R d\hat{\rho}_\lambda 
	\leq 
	\mathcal{C'} \frac{ 
s^* \log \left( 
\frac{n\sqrt{d} }{ s^*}	\right)
		+ \log\left(1/\epsilon\right)
	}{n} 
	,
	\end{align*}
	where $ \mathcal{C'}:= \mathcal{C}_{1,C_1,C',C_{\rm x}}  $ is a universal constant depending only on $ C_1,C',C_{\rm x} $.
\end{corollary}

\begin{remark}
According to Corollary \ref{corola_for_noiselesscase}, the bound on the misclassification excess risk for our proposed method follows an order of $ s^*\log(de/s^*)/n $. Meanwhile, under Assumption \ref{assume_margin}, the study in \cite{abramovich2018high} established a minimax lower bound for the misclassification excess risk, which is of the order $ s^*\log(de/s^*)/n$. Notably, this lower bound is also achieved by the logistic Slope estimator in that same paper under additional Weighted Restricted Eigenvalue condition. As a result, in the noiseless case, our rate is demonstrated to be minimax-optimal in a setting that $ n\leq e\sqrt{d} $, by noting that $ \frac{n\sqrt{d} }{ s^*} = \frac{n }{ e\sqrt{d}} \frac{de }{ s^*} $.
\end{remark}

In analogy to Theorem \ref{thrm_contraction_slow}, given  additional Assumption \ref{assume_margin}, we can establish that a stochastic classifier, $ \hat{\beta} \sim \hat{\rho}_{\lambda} $, drawn from our proposed pseudo-posterior in Equation \eqref{eq_mainporsterior} exhibits a fast rate. This particular finding is sometimes denominated as the contraction rate of the pseudo-posterior. 
\begin{theorem}
	\label{thrm_contraction} 
	Under the same assumptions for Theorem~\ref{thm_fastrate}, and the same definition for $\tau$ and $\lambda $, let $\varepsilon_n $  be any sequence in $(0,1) $ such that $\varepsilon_n\rightarrow 0 $ when $n\rightarrow\infty$. Define
	\begin{align*}
	\Omega_n 	= 
	\Biggl\{ \beta \in \mathbb{R}^{d}: 
	 R 
	 \leq 
	 (1+3C') R^*  
+
\mathcal{C}_{C,C_1,C',C_{\rm x}}  \frac{ 
	s^* \log ( n\sqrt{d}/ s^* )
	+ \log\left(1/\varepsilon_n \right)
}{n} 
	 \Biggr\}
	 .
	\end{align*}
	Then
	$$ \mathbb{E} \Bigl[ \mathbb{P}_{ \hat{\beta} \sim \hat{\rho}_{\lambda}} 
	( 	\hat{\beta} \in \Omega_n 	) \Bigr] 
	\geq 
	1 - 2 \varepsilon_n \xrightarrow[n\rightarrow \infty]{} 1.  
	$$
\end{theorem}

The results presented in Theorem \ref{thrm_contraction_slow} and Theorem \ref{thrm_contraction} are also novel findings, to the best of our knowledge.

\begin{remark}
In this section, we demonstrate the existence of specific values for the tuning parameters $ \lambda $ and $ \tau $ in our proposed method in theoretical results for prediction errors. It is important to acknowledge, however, that these values may not be the most suitable for practical applications. In practical applications, cross-validation can be employed to appropriately fine-tune these parameters. Nevertheless, the theoretical values identified in our analysis provide valuable insights into the expected magnitude of these tuning parameters when applied in practical situations.
\end{remark}

\subsection{Sharp rates with known sparsity $ s^* $}
\label{sc_known_sstar}

In the present section, we operate under the assumption that $ s^* $, denoting the number of non-zero coefficients in $ \beta^* $, is a known quantity. This assumption allows us to obtain results that sharply align with the rates established in \cite{abramovich2018high} (in the noiseless case). The first refinement is observed in the context of Theorem \ref{thm_slowrate}, where the tuning parameters $ \lambda $ and $ \tau $ are intricately dependent on the specific value of $ s^* $.
\begin{Proposition}
	\label{thm_slowrate_known_sstars}
	Assume that Assumption \ref{assume_bound_on_X}, \ref{assume_bound_on_thetruebayes} hold. We have, for  $ \tau = s^*/( n\sqrt{d} ) $ and $ \lambda = \sqrt{n s^* \log (de/s^*) } $, with probability at least $ 1-2\epsilon, \epsilon\in(0,1) $, that for all $ \beta^* $ such that $ \|\beta^*\|_1 \leq C_1-2d\tau $ and $ \|\beta^*\|_0 \leq s^* $,
	\begin{align*}
	\int R d\hat{\rho}_\lambda 
	\leq 
	(1+2C')R^*  
	+
	c \frac{\sqrt{s^* \log \left(de/s^*\right) }  }{\sqrt{n}} 
	+ 
	\frac{\log\left(1/\epsilon\right)}{\sqrt{n s^* \log (de/s^* ) }} 
	,
	\end{align*}
	where $ c $ is a universal constant depending only on $ C_1,C',C_{\rm x} $.
\end{Proposition}

\begin{remark}
	The proof of Proposition \ref{thm_slowrate_known_sstars} is given in Appendix \ref{sc_proof_knownstars}. In the noiseless case, i.e. $ R^* =0 $, the rate in Proposition \ref{thm_slowrate_known_sstars} is exactly matched the result from Theorem 2 in \cite{abramovich2018high} that proved a lower bound of order $ \sqrt{s^*\log(de/s^*)/n} $.
\end{remark}

The subsequent refinement becomes apparent within the framework of Theorem \ref{thm_fastrate}, wherein the dependency on the value of $ s^* $ is now explicitly confined to the tuning parameter $ \tau $, while other parameters remain unaffected.
\begin{Proposition}
	\label{thm_fastrate_known_sstars}
	Assume that Assumption \ref{assume_bound_on_X}, \ref{assume_bound_on_thetruebayes}, \ref{assume_margin} hold. We have, for $ \lambda = 2 n/(3C + 2) $, $ \tau =  s^*/(n\sqrt{d}) $ and with probability at least $ 1-2\epsilon, \epsilon\in(0,1) $, that  for all $ \beta^* $ such that $ \|\beta^*\|_1 \leq C_1-2d\tau $ and $ \|\beta^*\|_0 \leq s^* $,
	\begin{align*}
	\int R d\hat{\rho}_\lambda 
	\leq 
(1+3C') R^*  
	+
	\mathcal{C}_{C,C_1,C', C_{\rm x}} \frac{ 
		s^* \log ( de/ s^* )
		+ \log\left(1/\epsilon\right)
	}{n} 
	,
	\end{align*}
	where $ \mathcal{C}_{C,C_1,C', C_{\rm x}}  $ is a universal constant depending only on $ C,C_1,C', C_{\rm x} $.
\end{Proposition}

\begin{remark}
	The proof of Proposition \ref{thm_fastrate_known_sstars} is given in Appendix \ref{sc_proof_knownstars}. In the noiseless case, i.e. $ R^* =0 $, the rate in Proposition \ref{thm_fastrate_known_sstars} is exactly matched the result from Theorem 4 in \cite{abramovich2018high}, which is $ s^*\log(de/s^*)/n $.
\end{remark}

\begin{remark}
It is emphasized that all the oracle inequalities presented in the paper are not sharp; in other words, instead of $ R(\hat{\beta}) \leq R^* + \ldots $, we establish $ R(\hat{\beta}) \leq (1+\delta)R^* + \ldots $ with $ \delta>0 $. Therefore, the pursuit of achieving sharp oracle inequalities for our method remains a significant unresolved issue, which we defer to future investigations.
\end{remark}

\section{Numerical studies}
\label{sc_simulation}

\subsection{Implementation and comparison of methods}
\subsubsection*{Implementation}
In this section, we introduce the use of the Langevin Monte Carlo (LMC) algorithm as a method for sampling from the pseudo-posterior. 
The LMC algorithm is a gradient-based method for sampling from a  distribution. 

First, a constant step-size unadjusted LMC algorithm, as described in \cite{durmus2019high}, is proposed. The algorithm starts with an initial matrix $\beta_0 $ and uses the recursion:
\begin{equation}
\label{langevinMC}
\beta_{s+1} = \beta_{s} - h\nabla \log \hat{\rho}_{\lambda}(\beta_s) +\sqrt{2h} E_s\qquad s=0,1,\ldots
\end{equation}
where $h>0$ is the step-size and $ E_0, E_1,\ldots$ are independent random vectors with i.i.d standard Gaussian entries. It is essential to exercise caution when selecting the step size $h$, as an insufficiently small value may lead to the summation exploding, as highlighted in \cite{roberts2002langevin}. As an alternative method to ensure convergence to the desired distribution, one can incorporate a Metropolis–Hastings (MH) correction into the algorithm. However, this approach tends to slow down the algorithm due to the additional acceptance/rejection step required at each iteration.

The updating rule presented in \eqref{langevinMC} is now regarded as a proposal for a new candidate.
\begin{align*}
\tilde{\beta}_{s+1} = \beta_{s} - h\nabla \log \hat{\rho}_{\lambda} (\beta_s) +\sqrt{2h} E_s,\qquad
s=0,1,\ldots,
\end{align*}
This proposal is accepted or rejected in accordance with the Metropolis–Hastings algorithm, with the following probability:
$$
\min \left\lbrace 1, \frac{ \hat{\rho}_{\lambda} (\tilde{\beta}_{s+1}) q(\beta_s | \tilde{\beta}_{s+1}) }{
	\hat{\rho}_{\lambda} (\beta_s ) q(\tilde{\beta}_{s+1} | \beta_s ) } \right\rbrace,
$$
where
$
q(x' | x) 
\propto 
\exp \left(- \| x'-x + h\nabla \log \hat{\rho}_{\lambda} (x) \|^2 / (4h) \right)
$
is the transition probability density from $x$ to $x'$. This is recognized as the Metropolis-adjusted Langevin algorithm (MALA), ensuring convergence to the (pseudo) posterior. In contrast to the random-walk Metropolis-Hastings (MH), MALA typically suggests moves towards regions with higher probability, enhancing the likelihood of acceptance. The selection of the step-size $h$ for MALA aims to achieve an acceptance rate of approximately 0.5, as recommended by \cite{roberts1998optimal}. In the same configuration, the step-size for Langevin Monte Carlo (LMC) is chosen to be smaller than those for MALA.

\subsubsection*{Comparison of Methods}
We will evaluate the efficacy, in term of prediction error, of our proposed methodologies by comparing them to Bayesian approaches that utilize logistic regression, as elucidated in Section \ref{sc_modelstatement}. In this scenario, the pseudo-likelihood $ \exp(-\lambda r^\ell (M)) $, with $ \lambda = n $, aligns precisely with the likelihood of the logistic model. Here, 
\[ r^\ell (M) = \frac{1}{n}\sum_{i=1}^N \text{logit}\left(Y_i (\beta^\top x_i)\right), \]
where \( \text{logit}(u) = \log(1+e^{-u}) \) represents the logistic loss. It is important to note that the prior distribution remains consistent with the previous sections. As investigated in \cite{zhang2004statistical}, the logistic loss can function as a convex substitute for the hinge loss, providing an approximation to the 0-1 loss. However, it is essential to highlight that utilizing the logistic loss may result in a slower convergence rate compared to the hinge loss, as discussed in \cite{zhang2004statistical}.

In this investigation, we assess the effectiveness of our suggested approaches using hinge loss, identified as H$_{_{\rm LMC}}$ and H$_{_{\rm MALA}}$ for the LMC and MALA algorithms, respectively. We compare these methods with three other alternatives: (1) Logit$_{_{\rm LMC}}$, (2) Logit$_{_{\rm MALA}}$, both based on Bayesian logistic regression, and (3) the logistic Lasso, which represents a contemporary and highly-regarded method. The logistic Lasso is a frequentist technique, and its implementation is available in the \texttt{R} package ``\texttt{glmnet}" \cite{glmnet}.

\begin{table}[!h]
	\centering
	\caption{Outline of simulation settings.}
	\begin{tabular}{ | l llll | }
		\toprule
		Setting & Name &  $Z $ & $N $ &
		\\
		\hline
		I.1 & Hinge 	& $Z =1$ & $N=0 $ & 
		\\
		I.2 & Hinge with noise & $Z =1 $ & $N\sim \mathcal{N}(0,1) $  & 
		\\
		I.3 & Hinge with switch 	& $ Z\sim 0.9 \delta_1 + 0.1 \delta_{-1}$ & $N=0$  & 
		\\
		I.4 & Hinge with switch and noise 	& $Z\sim 0.9 \delta_1 + 0.1 \delta_{-1}$ & $N\sim \mathcal{N}(0,1) $  & 
		\\
		II.1 & logistic 	& $Z =1 $ & $N=0 $ & 
		\\
		II.2 & logistic with switch	& $ Z\sim 0.9 \delta_1 + 0.1 \delta_{-1}$ & $N=0 $ & 
		\\
		II.3 & logistic with noise & $Z =1 $ & $ N\sim \mathcal{N}(0,1) $ & 
		\\
		II.4 & logistic with noise and switch & $ Z\sim 0.9 \delta_1 + 0.1 \delta_{-1}$ & $N\sim \mathcal{N}(0,1) $ & 
		\\
		\bottomrule
	\end{tabular}
	\label{tb_type_noise}
\end{table}

\subsection{Simulation setup}
We examine various scenarios for data generation to evaluate the performance of our approach. Initially, we consider a small-scale setup with dimensions \(n=50, p=100\). In this initial configuration, the sparsity, or the number of non-zero coefficients in the true parameter \(\beta^*\), is set as \(s_0=10\). Subsequently, we explore a larger setup with \(n=200, p=1000\). In this second configuration, the sparsity of the true parameter \(\beta^*\) is adjusted between \(s_0=100\) and \(s_0=10\), with the latter denoting a highly sparse model. The entries of the covariate matrix \(X\) are generated from a normal distribution \(\mathcal{N}(0,1)\). In all instances, the non-zero coefficients of \(\beta^*\) are independently and identically drawn from \(\mathcal{N}(0,10^2)\).

Next, we explore the following settings to obtain the responses:
\begin{itemize}
	\item Setting I: 
	$$ Y = {\rm sign}( X \beta^* + N ) Z. $$
	\item Setting II: with 
	$
	u = X\beta^* + N, 
	$ 
	put $ p= 1/ (1 + e^{-u} ) $: 
	$$ Y_{i} \sim  \text{Binomial}(p_{i})Z.
	$$	
\end{itemize}
In this context, the variability in the noise term \( (N,Z) \) results in distinct scenarios, each contributing to a different setup in every setting. A summary of these variations is provided in Table \ref{tb_type_noise}.

The LMC, MALA are run with 30000 iterations and  the first 5000 steps are discarded as burn-in period. The LMC is initialized at the logistic Lasso while the MALA is initialized at zero-vector. We set the values of tuning parameters $\lambda$ and $\tau$ to 1 for all scenarios. It is important to acknowledge that a better approach could be to tune these parameters using cross validation, which could lead to improved results. The logistic Lasso method is run with default options and that 10-fold cross validation is used to select the tuning parameter.

Each simulation setting is repeated 100 times and we report the averaged results for the misclassification rate. The results of the simulations study are detailed in Table \ref{tb_small_s10}, Table \ref{tb_large_s100}, Table \ref{tb_Lag_s10_veryspase}  and the values within parentheses indicate the standard deviations associated with each misclassification rate percentage.

\begin{table}[htb]
	\centering
	\caption{Misclassification rate. $ n=50, p =100,s_0=10 $.}
	\begin{tabular}{ | l lllll | }
		\hline\hline
		Setting & Logit$_{_{\rm LMC}}$(\%) & H$_{_{\rm LMC}}$(\%) & Logit$_{_{\rm MALA}}$(\%) & H$_{_{\rm MALA}}$ (\%) & Lasso (\%)
		\\ 
		\hline
		I.1 & 4.30 (2.93) & 1.36 (1.63) & 6.74 (3.80) & 2.92 (2.30) & 5.76 (7.38)
		\\
		I.2 & 4.36 (3.20) & 1.26 (1.52) & 6.40 (3.67) & 2.38 (2.14) & 7.02 (6.75) 
		\\
		I.3 & 13.9 (5.54) & 11.1 (4.59) & 15.0 (5.18) & 12.0 (4.90) & 15.6 (8.09) 
		\\
		I.4 & 13.7 (5.44) & 11.4 (5.17) & 15.9 (5.68) & 12.0 (4.97) & 16.3 (12.1) 
		\\ 
		\hline
		II.1 & 2.16 (2.60) & 0.66 (1.14) & 5.82 (3.71) & 2.42 (2.47) & 3.60 (11.1) 
		\\
		II.2 & 2.08 (2.14) & 0.54 (0.98) & 6.40 (3.49) & 3.08 (2.21) & 6.58 (7.00) 
		\\
		II.3 & 11.2 (4.27) & 10.0 (4.09) & 14.3 (5.05) & 11.6 (4.10) & 14.6 (7.90) 
		\\
		II.4 & 12.0 (4.89) & 10.3 (4.17) & 15.2 (5.56) & 12.5 (4.90) & 17.7 (14.5) 
		\\
		\hline
	\end{tabular}
\label{tb_small_s10}

	\caption{Misclassification rate. $ n=200, p =1000,s_0=100 $.}
	\begin{tabular}{ | l lllll | }
		\hline\hline
		Setting & Logit$_{_{\rm LMC}}$ (\%) & H$_{_{\rm LMC}}$(\%) & Logit$_{_{\rm MALA}}$ (\%) & H$_{_{\rm MALA}}$ (\%) & Lasso (\%)
		\\ 
		\hline
		I.1 & 4.60 (1.99) & 1.26 (0.91) & 7.52 (2.29) & 3.04 (1.22) & 8.21 (11.4) 
		\\
		I.2 & 4.76 (2.20) & 1.40 (0.90) & 7.93 (2.67) & 3.00 (1.18) & 12.4 (12.2) 
		\\
		I.3 & 14.1 (2.84) & 10.8 (2.14) & 16.2 (3.10) & 12.2 (2.23) & 21.8 (11.2) 
		\\
		I.4 & 14.1 (262) & 10.9 (2.10) & 16.4 (2.50) & 12.7 (2.08) & 20.5 (15.2) 
		\\ \hline
		II.1 & 4.97 (1.91) & 1.27 (0.82) & 7.90 (2.68) & 3.02 (1.33) & 9.26 (10.0) 
		\\
		II.2 & 5.19 (2.12) & 1.26 (0.79) & 7.48 (2.44) & 2.96 (1.25) & 10.2 (11.9) 
		\\
		II.3 & 13.8 (2.56) & 11.1 (2.34) & 16.5 (3.01) & 12.7 (2.41) & 20.3 (11.2) 
		\\
		II.4 & 14.3 (3.26) & 11.3 (2.50) & 16.2 (3.63) & 12.7 (2.43) & 19.7 (10.7) 
		\\
		\hline
	\end{tabular}
\label{tb_large_s100}

	\caption{Misclassification rate. $ n=200, p =1000,s_0=10 $.}
	\begin{tabular}{ | l lllll | }
		\hline\hline
		Setting & Logit$_{_{\rm LMC}}$ (\%) & H$_{_{\rm LMC}}$(\%) & Logit$_{_{\rm MALA}}$ (\%) & H$_{_{\rm MALA}}$ (\%) & Lasso (\%)
		\\ 
		\hline
		I.1 & 4.50 (1.68) & 1.22 (0.81) & 7.20 (2.21) & 2.91 (1.22) & 4.25 (3.56) 
		\\
		I.2 & 4.34 (1.71) & 1.20 (0.79) & 7.34 (2.25) & 3.11 (1.21) & 3.84 (3.35) 
		\\
		I.3 & 14.0 (2.34) & 10.7 (2.36) & 16.0 (2.98) & 11.8 (2.48) & 12.7 (4.41) 
		\\
		I.4 & 14.1 (2.62) & 10.9 (2.46) & 16.4 (2.97) & 12.4 (2.36) & 12.8 (4.77) 
		\\ \hline
		II.1 & 4.16 (1.68) & 1.04 (0.77) & 7.13 (2.36) & 2.78 (1.17) & 3.75 (3.51) 
		\\
		II.2 & 4.33 (1.47) & 1.10 (0.83) & 7.12 (2.02) & 2.82 (1.16) & 4.04 (3.74) 
		\\
		II.3 & 14.6 (2.91) & 11.1 (2.31) & 16.9 (2.59) & 12.4 (2.24) & 13.2 (4.74) 
		\\
		II.4 & 14.7 (2.77) & 11.2 (2.26) & 16.8 (2.71) & 12.7 (2.33) & 13.1 (4.96) 
		\\
		\hline
	\end{tabular}
	\label{tb_Lag_s10_veryspase}
\end{table}

\subsection{Results from simulations}

The exhaustive analysis of results extracted from Table \ref{tb_small_s10}, Table \ref{tb_large_s100}, and Table \ref{tb_Lag_s10_veryspase} provides a thorough insight into the robust performance of our proposed methods when compared to the Lasso and Bayesian logistic approaches. Significantly, throughout all simulated scenarios, the H$_{_{\rm LMC}}$ method, implemented via the Langevin Monte Carlo (LMC) algorithm, consistently showcases the smallest misclassification rate. This substantiates the efficacy  of our approach across a diverse array of settings, spanning variations in sample size, sparsity levels, and distinct noise settings.

A noteworthy aspect is the exceptional performance highlighted in Table \ref{tb_small_s10}, where H$_{_{\rm LMC}}$ outperforms the Lasso method by nearly tenfold. This pronounced superiority is particularly evident in scenarios characterized by small sample sizes, reinforcing the robustness of our proposed H$_{_{\rm LMC}}$ method under challenging conditions. This outcome underscores the method's ability to navigate challenges related to limited data, emphasizing its potential applicability in practice where small sample sizes are prevalent.

The second most effective strategy emerges from our proposed method implemented using the Metropolis-adjusted Langevin Algorithm (MALA), referred to as H$_{_{\rm MALA}}$. It consistently secures the second-best position across various scenarios. Notably, in almost all cases, except for the logistic model scenario in Table \ref{tb_small_s10}, where Logit$_{_{\rm LMC}}$ exhibits a slight advantage, H$_{_{\rm MALA}}$ stands out as the runner-up. Even in scenarios where other methods show slight advantages, H$_{_{\rm LMC}}$maintains dominance. Comparing Logit$_{_{\rm MALA}}$ with the logistic Lasso reveals nuanced results. In less sparse situations, exemplified by Table \ref{tb_large_s100}, Logit$_{_{\rm MALA}}$ demonstrates a slight performance edge. However, in the case of highly sparse models, as exemplified in Table \ref{tb_Lag_s10_veryspase}, the logistic Lasso emerges as the more efficient choice. 

Generally, as observed from Table \ref{tb_large_s100} to Table \ref{tb_Lag_s10_veryspase}, there is a tendency for all considered methods to reduce the misclassification rate as the sparsity level increases. This trend is particularly notable for the logistic Lasso. However, the enhancements in performance for our methods (H$_{_{\rm LMC}}$ and H$_{_{\rm MALA}}$) are modest, indicating their adaptability to varying levels of sparsity. Similarly, with an increase in both dimension and sample size, as demonstrated from Table \ref{tb_small_s10} to Table \ref{tb_Lag_s10_veryspase}, there are also noticeable performance improvements in our methods (H$_{_{\rm LMC}}$ and H$_{_{\rm MALA}}$), as well as the logistic Lasso. These findings offer valuable insights into the strengths and limitations of each method, facilitating informed decision-making based on the specific characteristics of the dataset under consideration.

\subsection{An application: Prostate  tumor classification with microarray gene expression data}
\label{sc_realdata}
In this section, we evaluate the performance of our proposed methods on a real data.

\begin{table}[htb]
	\centering
	\caption{Misclassification rate for real prostate data.}
	\begin{tabular}{ |  lllll | }
		\hline\hline
		Logit$_{_{\rm LMC}}$ (\%) & H$_{_{\rm LMC}}$(\%) & Logit$_{_{\rm MALA}}$ (\%) & H$_{_{\rm MALA}}$ (\%) & Lasso (\%)
		\\ 
		\hline
		9.71 (4.41) & 9.58 (4.31) & 9.74 (4.47) & 9.55 (4.42) & 9.77 (4.31) 
		\\
		\hline
	\end{tabular}
	\label{tb_real_data}
\end{table}

The ``prostate" dataset is accessible through the \texttt{R} package ``\texttt{spls}" \cite{pkgspls}. It comprises 52 samples corresponding to prostate tumors and 50 samples corresponding to normal tissue. The response variable $ Y $ encodes normal and tumor classes as 0 and 1, respectively. The covariates matrix $ X $, with dimensions 102 rows by 6033 columns, represents gene expression data. Preprocessing steps, including normalization, log transformation, and standardization to achieve zero mean and unit variance across genes, were applied to the arrays, following the procedures outlined in \cite{dettling2004bagboosting,dettling2002supervised}. Additional details can be found in \cite{chung2010sparse}.

The dataset is randomly partitioned into two subsets: a training set comprising $ 71 $ samples and a test set comprising $ 31 $ samples, roughly representing $ 70/30 $ percent of the total samples. The training data is utilized for running the methods, and their prediction accuracy is assessed based on the test data. This process is repeated 100 times, each instance involving a distinct random partition of the training and test data. The outcomes of this iterative procedure are depicted in Table \ref{tb_real_data}. This strategy enables us to accommodate potential fluctuations in the data and gain a more thorough comprehension of the methods' performance.

The findings presented in Table \ref{tb_real_data} indicate that all the methods under consideration exhibit effective performance, yielding comparable results. Specifically, H$_{_{\rm MALA}}$ demonstrates an error rate of 9.55\%, showcasing proficiency; however, this improvement is marginal when compared to the 9.77\% error rate achieved by Lasso. The similarity in the outcomes suggests that, in this context, the performance distinctions between H$_{_{\rm MALA}}$ and Lasso are relatively small.

\section{Discussion and Conclusion}
\label{sc_disandconclusion}
In this work, we present an innovative probabilistic framework designed to address the challenges associated with high-dimensional sparse classification problems. Our approach involves the utilization of exponential weights associated with the empirical hinge loss, leading to the establishment of a pseudo-posterior distribution within a class of sparse linear classifiers. Notably, we introduce a sparsity-inducing prior distribution over this class, utilizing a scaled Student's t-distribution with 3 degrees of freedom.

By employing the PAC-Bayesian bound technique, we derive comprehensive theoretical insights into our proposed methodology, particularly focusing on prediction errors. Specifically, under the low-noise condition, we demonstrate that our approach exhibits a fast rate of convergence of order $ n^{-1} $. Importantly, in the noiseless case, our analysis reveals that the prediction error achieved is minimax-optimal. Furthermore, we establish the contraction rate of our pseudo-posterior, presenting novel findings in the current literature.

Beyond the robust theoretical foundation, our approach facilitates practical implementation insights. We leverage the Langevin Monte Carlo method, a gradient-based sampling approach, to demonstrate the applicability of our framework. Through extensive simulations and a real data application, our method showcases enhanced robustness across various scenarios, such as varying sample sizes and sparsity levels. Numerical results highlight the superior performance of our approach compared to the logistic Lasso, a widely recognized state-of-the-art method.

Looking ahead, future investigations could delve into the estimation challenges posed by our methodology. Additionally, a crucial aspect not addressed in this paper pertains to variable selection, a topic of paramount importance in practical applications. This opens avenues for further research and exploration in enhancing the versatility and applicability of our proposed approach.

\subsection*{Acknowledgments}
The author would like to thank the Editor, Associate Editor and the reviewer who kindly reviewed the earlier version of this paper and provided valuable suggestions and enlightening comments. The author is indebted to the Associate Editor for the helpful suggestions on references on heavy tailed priors, and to the reviewer for critical comments on the proof.
The author acknowledges support from the Norwegian Research Council, grant number 309960 through the Centre for Geophysical Forecasting at NTNU.

\subsection*{Conflicts of interest/Competing interests}
The author declares no potential conflict of interests.

\subsection*{Data and codes}
Data and Rcodes are available at 
\url{https://github.com/tienmt/sparseclassificationEWA/}.

\newpage
\appendix
\section{Proofs}
\label{sc_proof}

For any $\Theta\subset \mathbb{R}^{d}$,  let $\mathcal{P}(\Theta)$ denote the set of all probability distributions on $\Theta$ equipped with the Borel $\sigma$-algebra. 
For $(\mu,\nu)\in \mathcal{P}(\Theta)^2$, $\mathcal{K}(\nu,\mu) $ denotes the Kullback-Leibler divergence. The following Donsker and Varadhan's lemma is an important key to establish our results.

\begin{lemma}[\cite{catonibook}{Lemma 1.1.3}]
	\label{lemma_donvara}
	
	Let $\mu \in\mathcal{P}(\Theta)$. For any measurable, bounded function $h:\Theta\rightarrow\mathbb{R}$ we have:
	\begin{equation*}
	\log \int {\rm e}^{h(\theta)} \mu({\rm d}\theta) = 
	\sup_{\rho\in\mathcal{P}(\Theta)}\left[\int h(\theta) \rho({\rm d}\theta) 
	-
	\mathcal{K} (\rho , \mu)\right].
	\end{equation*}
	Moreover, the supremum w.r.t $\rho$ in the right-hand side is
	reached for the Gibbs distribution,
$
\hat{\rho} (d\theta)
\propto 
\exp(h(\theta)) \pi (d\theta).
$
\end{lemma}

\subsection{Proof for slow rate}

We remind that $ R^* = R(\beta^*), r_n^* = r_n(\beta^*) $. We remind here a version of Hoeffding's inequality
for bounded random variables.
\begin{lemma}
	\label{lemma_hoeffding}
	Let $ U_i, i = 1,\ldots,n$ be $n$ independent random
	variables with $ a \leq U_i \leq b $ a.s., and
	$\mathbb{E}(U_i)=0$. Then, for any $\lambda>0$,
	$$ 
	\mathbb{E}\exp\left(\frac{\lambda}{n}
	\sum_{i=1}^{n} U_i\right)
	\leq
	\exp\left(\frac{\lambda^2 (b-a)^2}{8n} \right) .$$
\end{lemma}

\begin{proof}[\bf Proof of Theorem~\ref{thm_slowrate}]
		\text{}
	\\
	\textit{Step 1:}
	
	Put
	$$
	U_{i} 
	=  
	\mathbbm{1}_{Y_{i} ( \beta^\top x_{i} ) \leq 0} - 
	\mathbbm{1}_{Y_{i} ( \beta^{*\top}  x_{i} ) \leq 0} 
	.
	$$
Then, $ -1 \leq U_i \leq 1 $ a.s., we  apply the Hoeffding's Lemma \ref{lemma_hoeffding} to get
		\begin{align*}
	 \mathbb{E}\exp\left\{\lambda[R( \beta )-R^* ]-\lambda[
	r_n( \beta )-r_n^* ] 
	\right\}
	\leq
	\exp\left\{
	\frac{\lambda^2 }{2n} 
	\right\}
	.
	\end{align*}
	We obtain, for any $\lambda\in(0,n) $, 
	\begin{align*}
	\int \mathbb{E}\exp\left\{\lambda[R( \beta )-R^* ]-\lambda[
	r_n( \beta )-r_n^* ] 
	-
	\frac{\lambda^2 }{2n}
	\right\} 
	{\rm d}\pi( \beta ) 
	\leq 
	1
	,
	\end{align*}
	and, using the Fubini's theorem, we  get that
	\begin{align}
	\label{eq_t_contraction_slow}
 \mathbb{E} \int \exp\left\{\lambda[R( \beta )-R^* ]-\lambda[
r_n( \beta )-r_n^* ] 
-
\frac{\lambda^2 }{2n} 
\right\} 
{\rm d}\pi( \beta ) 
\leq 
1
,
\end{align}	
	Consequently, using Lemma \ref{lemma_donvara},
\begin{align*}
\mathbb{E} \exp \left\lbrace \sup_{\rho} \int 
\left\{ 
\lambda [R( \beta )-R^* ] 
-
\lambda[r_n( \beta )-r_n^* ]
-
\frac{\lambda^2 }{2n} 
\right\}
\rho (d \beta) - 
\mathcal{K}(\rho,\pi) \right\rbrace 
\leq 1.
\end{align*}
Using Markov's inequality, for $ \epsilon \in (0,1) $,
\begin{align*}
\mathbb{P} \left( \sup_{\rho} \int 
\left\lbrace 
\lambda 
[R( \beta )-R^* ]-\lambda [r_n( \beta )-r_n^* ] 
-
\frac{\lambda^2 }{2n}  \right\rbrace
\rho (d\beta ) 
- \mathcal{K}(\rho,\pi) +\log \epsilon>0 \right)  \leq \epsilon.
\end{align*}
Then taking the complementary and we obtain  with probability at least $1-\epsilon$ that:
\begin{align*}
\forall \rho, \quad \lambda \int  
[R( \beta )-R^* ]\rho (d\beta ) 
\leq 
\lambda \int 
[r_n( \beta )-r_n^* ]  \rho (d\beta ) + \mathcal{K}(\rho,\pi) 
+
\frac{\lambda^2 }{2n} 
+ 
\log 
\frac{1}{\epsilon}.
\end{align*}
Now, note that as $ r^h_n \geq r_n $ and
as it stands for all $\rho$ then the right hand side can be minimized and, from Lemma \ref{lemma_donvara}, the minimizer over $\mathcal{P}(\mathbb{R}^{d}) $ is $\hat{\rho}_\lambda$.	Thus we get, when $\lambda > 0 $,
\begin{align*}
\int R d\hat{\rho}_\lambda 
\leq 
R^*  + 
\inf_{\rho \in \mathcal{P}(\mathbb{R}^{d}) } 
\left[ \int 
r^h_nd\rho + \frac{1}{\lambda} \mathcal{K}(\rho,\pi) 
\right] 
- 
r_n^*  
+
\frac{\lambda }{2n} 
+ 
\frac{1}{\lambda} 
\log\frac{1}{\epsilon}
.
\end{align*}	

	\textit{Step 2:}
	\\ 
First, we have that,
	\begin{align}
	\int r^h_n (\beta) \rho (d\beta) 
	&	=  	
	\frac{1}{n} \int \sum_{i=1}^n ( 1 - Y_{i} (\beta^\top x_i ) )_{_+} \,  \rho (d\beta) 
	\nonumber
	\\
	& \leq
	\frac{1}{n} \left[ \sum_{i=1}^n 
	( 1 -  Y_{i} (\beta^{*\top} x_i ))_+
	+
	\int \sum_{i=1}^n  
	\left| 
	( \beta - \beta^{*} )^\top x_i  
	\right|  \rho (d\beta)  \right]
	\nonumber
	\\
	& \leq
	r^h_n(\beta^{*}) 
	+ 	
	\frac{1}{n} \sum_{i=1}^n 	\int 
	\|  \beta - \beta^{*}\|_2 \|x_i\|_2
	\rho (d\beta)
	\nonumber
	\\
	& \leq
	r^h_n(\beta^{*}) 
	+ 	
	C_{\rm x}	\int 
	\|  \beta - \beta^{*}\|_2 
	\rho (d\beta)
	.
	\label{eq_boundforhinge}
	\end{align}
	And for $ \rho = p_0  $, as in \eqref{eq_specific_distribution}, and using Lemma \ref{lema_boundfor_ell2},
	\begin{align*}
	\int 
	\|  \beta - \beta^{*}\|_2 
	p_0 (d\beta)
	\leq 
	\left(	\int 
	\|  \beta - \beta^{*}\|_2^2 
	p_0 (d\beta) \right)^{1/2}
	\leq 
	2 \tau \sqrt{d}
	. 
	\end{align*}
	From Lemma \ref{lema_boundof_KL}, we have that
	$$
	\mathcal{K}( p_0 ,  \pi) 
	\leq 
	4 s^* \log \left(\frac{C_1 }{\tau s^*}\right)
	+
	\log(2)
	.
	$$
From Assumption \ref{assume_bound_on_thetruebayes}, we have   $r^h_n(\beta^*) \leq (1+C')r^*_n $, we obtain
	\begin{align*}
	\int R d\hat{\rho}_\lambda 
	\leq 
	R^* + 	C'r^*_n
	+
	C_{\rm x} 2 \tau \sqrt{d}
	+ 
	\frac{	4s^* \log \left(\frac{C_1 }{\tau s^*}\right)
		+
		\log(2) }{\lambda} 
	+
	\frac{\lambda }{2n} 
	+
	\frac{1}{\lambda} \log\left(\frac{1}{\epsilon}\right) 
	.
	\end{align*}
	Then, we use Lemma \ref{lem_theo1}, with probability at least $ 1-2\epsilon $, to obtain that
	\begin{align}
	\label{eq_used_forknwon_sstatr}
\int R d\hat{\rho}_\lambda 
\leq 
(1+2C')R^*  
+
C'\frac{1}{n\varsigma }\log \frac{1}{\epsilon}
+
C_{\rm x} 2 \tau \sqrt{d}
+ 
\frac{	4s^* \log \left(\frac{C_1 }{\tau s^*}\right)
	+
	\log(2) }{\lambda} 
+
\frac{\lambda }{2n} 
+
\frac{1}{\lambda} \log\left(\frac{1}{\epsilon}\right) 
.
\end{align}
By taking  $ \tau = 1/( n\sqrt{d} ) $, we obtain that 
\begin{align*}
	\int R d\hat{\rho}_\lambda 
	\leq 
	(1+2C')R^*  
	+
C'	\frac{1}{n\varsigma }\log \frac{1}{\epsilon}
	+
	C_{\rm x} 
	\frac{	4s^* \log \left(\frac{n\sqrt{d}C_1 }{ s^*}\right)
		+
		\log(2) }{\lambda} 
+
\frac{\lambda }{2n} 
+
\frac{1}{\lambda} \log\left(\frac{1}{\epsilon}\right) 
.
\end{align*}

By taking $ \lambda = \sqrt{n \log (nd) } $, we can obtain that
\begin{align*}
\int R d\hat{\rho}_\lambda 
\leq 
(1+2C')R^*  
+  	C_{\rm x} 
\frac{	4s^* \log \left(\frac{n\sqrt{d} C_1 }{ s^*}\right)
 }{\sqrt{n \log (nd) } } 
+
\frac{\sqrt{ \log (nd) }}{2\sqrt{n}} 
+
 \left( \frac{1}{\sqrt{n \log (nd) }} + \frac{C'}{n\varsigma } \right) \log\left(1/\epsilon\right) 
.
\end{align*}

Therefore, we can obtain that
	\begin{align*}
\int R d\hat{\rho}_\lambda 
\leq 
(1+2C')R^*  
+
c \frac{s^* \sqrt{\log \left(n\sqrt{d}/s^*\right) }  }{\sqrt{n}} 
+ 
c\frac{\log\left(1/\epsilon\right)}{\sqrt{n \log (nd) }} 
,
\end{align*}
where $ c $ is a universal constant depending only on $ C', C_1, C_{\rm x} $.
 The proof is completed.
 
\end{proof}

\begin{proof}[\textbf{Proof of Theorem \ref{thrm_contraction_slow}}]
 Let $\varepsilon_n $  be any sequence in $(0,1) $ such that $\varepsilon_n\rightarrow 0 $ when $n\rightarrow\infty$.	From \eqref{eq_t_contraction_slow}, we have that
	\begin{align*}
	\mathbb{E} \Biggl[ \int \exp \left\{ 
	\lambda[R( \beta )-R^* ] 
	-\lambda[r_n( \beta )-r_n^* ]
	- 
	\log \left[\frac{d\hat{\rho}_{\lambda}}{d \pi} (\beta)  \right]
	-
	\frac{\lambda^2 }{2n} 
	- 
	\log\frac{1}{\varepsilon_n}
	\right\}
	\hat{\rho}_{\lambda}(d \beta)
	\Biggr]
	\leq 
	\varepsilon_n
	.
	\end{align*}
We now use the Chernoff's trick, i.e. using $\exp(x) \geq 1_{\mathbb{R}_{+}}(x)$, this yields:
	$$
	\mathbb{E} \Bigl[ 
	\mathbb{P}_{\beta \sim \hat{\rho}_{\lambda}} 
	(\beta \in \Theta_n ) \Bigr]
	\geq 
	1- \varepsilon_n
	$$
	where
	$$
	\Theta_n
	= 
	\left\{\beta : 	
	\lambda	[R( \beta )-R^* ] -\lambda[r_n( \beta )-r_n^* ]   
	\leq      
	\log \left[\frac{d\hat{\rho}_{\lambda}}{d \pi} (\beta)  \right]
	+ 	\frac{\lambda^2 }{2n} 
	+ \log\frac{2}{\varepsilon_n} \right\}.
	$$
	Using the definition of $\hat{\rho}_\lambda $ and noting that as $ r_n \leq r^h_n $, for $ \beta \in \Theta_n $ we have
	\begin{align*}
	\lambda	[ R(\beta) - R^* ]
	&\leq  
	\lambda\Bigl( r(\beta) - r_n^* \Bigr)  +       \log \left[\frac{d\hat{\rho}_{\lambda}}{d \pi} (\beta)  \right]
		+ 	\frac{\lambda^2 }{2n} 
	+ \log\frac{2}{\varepsilon_n}
	\\
	&    \leq  
	\lambda\Bigl( r^h_n(\beta) - r_n^* \Bigr)  +       \log \left[\frac{d\hat{\rho}_{\lambda}}{d \pi} (\beta)  \right]
		+ 	\frac{\lambda^2 }{2n} 
	+ \log\frac{2}{\varepsilon_n}
	\\
	& \leq 
	- \log\int\exp\left[
	-\lambda r^h_n (\beta) \right]\pi({\rm d} \beta) - \lambda r_n^*
		+ 	\frac{\lambda^2 }{2n} 
	+ \log\frac{2}{\varepsilon_n}
	\\
	& = \lambda\Bigl( \int r^h_n (\beta) \hat{\rho}_{\lambda}({\rm d} \beta) - r_n^* \Bigr)  +    \mathcal{K}(\hat{\rho}_\lambda,\pi)
		+ 	\frac{\lambda^2 }{2n} 
	+ \log\frac{2}{\varepsilon_n}
	\\
	& = 
	\inf_{\rho} \left\{ 
	\lambda\Bigl( \int r^h_n (\beta) \rho({\rm d}\beta) - r_n^* \Bigr)  +    \mathcal{K}(\rho,\pi)
		+ 	\frac{\lambda^2 }{2n} 
	+ \log\frac{2}{\varepsilon_n} \right\}.
	\end{align*}
	We upper-bound the right-hand side exactly as Step 2 in the proof of Theorem~\ref{thm_slowrate} (with Lemma \ref{lem_theo1}). The result of the theorem is followed.
	
\end{proof}

\subsection{Proof for fast rate}

We will make use of the following version of the Bernstein's lemma taken from \cite[page 24]{MR2319879}.
\begin{lemma}
	\label{lemmemassart} Let $U_{1}$, \ldots, $U_{n}$ be independent real
	valued random variables. Let us assume that there are two constants
	$v$ and $w$ such that
	$
	\sum_{i=1}^{n} \mathbb{E}[U_{i}^{2}] \leq v 
	$
	and that for all integers $k\geq 3$,
	$
	\sum_{i=1}^{n} \mathbb{E}\left[(U_{i})_+^{k}\right] \leq v k!w^{k-2}/2. 
	$
	Then, for any $ \zeta \in (0,1/w)$,
	$$ 
	\mathbb{E}
	\exp\left[\zeta \sum_{i=1}^{n}\left[U_{i}-\mathbb{E}U_{i}\right]
	\right]
	\leq 
	\exp\left(\frac{v\zeta^{2}}{2(1-w\zeta )} \right) .
	$$
\end{lemma}

\begin{proof}[\bf Proof of Theorem~\ref{thm_fastrate}]
	\text{}
	\\
	\textit{Step 1:}
	
	Fix any $ \beta $ and put
	$$
	U_{i} 
	=  
	\mathbbm{1}_{Y_{i} ( \beta^\top x_{i} ) \leq 0} - 
	\mathbbm{1}_{Y_{i} ( \beta^{*\top}  x_{i} ) \leq 0} 
	.
	$$
	Under Assumption \ref{assume_margin}, we have that
	$
	\sum_{i} \mathbb{E}[U_{i}^{2}]
	\leq
	nC[R( \beta )-R^* ] .$
	Now, for any integer $k\geq 3$, as the 0-1 loss is bounded, we have that
	$$
	\sum_{i} \mathbb{E}\left[(U_{i})_+^{k}\right] \leq
	\sum_{i} \mathbb{E}\left[ |U_{i}|^{k-2} |U_{i}|^{2}\right]
	\leq
	\sum_{i} \mathbb{E}\left[ |U_{i}|^{2}\right].
	$$ 
	Thus, we can apply Lemma~\ref{lemmemassart} with  $v := 	nC[R( \beta )-R^* ] $, $w:=1 $ and $\zeta := \lambda/n $. We obtain, for any $\lambda\in(0,n) $,
	\begin{align*}
	 \mathbb{E}\exp\{\lambda (
	  [R( \beta )-R^* ]-[
	r_n( \beta )-r_n^* ] )
	\} 
	\leq 
	 \exp \left\{ \frac{C\lambda^2[R( \beta )-R^* ] }{2n (1-\lambda/n)} 	 
	  \right\},
	\end{align*}
	and
	\begin{align*}
	\int \mathbb{E}\exp\left\{\lambda[R( \beta )-R^* ]-\lambda[
	r_n( \beta )-r_n^* ] 
	-
	\frac{C\lambda^2[R( \beta )-R^* ] }{2n(1-\lambda/n)} 
	\right\} {\rm d}\pi( \beta ) 
	\leq 
	1.
	\end{align*}
	Them, using Fubini's theorem, we get:
	\begin{align}
	\label{eq_to_contraction}
	\mathbb{E} \int \exp \left\{ 
	(\lambda- \frac{C\lambda^2 }{2n(1-\lambda/n)} )[R( \beta )-R^* ] 
	-\lambda[r_n( \beta )-r_n^* ]
	\right\} \pi(d\beta ) 
	\leq 1.
	\end{align}
	Consequently, using Lemma \ref{lemma_donvara},
	\begin{align*}
	\mathbb{E} \exp \left\lbrace \sup_{\rho} \int \left\{ 
	(\lambda- \frac{C\lambda^2 }{2n(1-\lambda/n)} )[R( \beta )-R^* ] 
	-\lambda[r_n( \beta )-r_n^* ]
	\right\}
	\rho (d M) - 
	\mathcal{K}(\rho,\pi) \right\rbrace 
	\leq 1.
	\end{align*}
	Using Markov's inequality,
	\begin{align*}
	\mathbb{P} \left( \sup_{\rho} \int 
	\left\lbrace 
	(\lambda-\frac{C\lambda^2 }{2n(1-\lambda/n)}) 
	[R( \beta )-R^* ]-\lambda [r_n( \beta )-r_n^* ]  \right\rbrace
	\rho (d\beta ) 
	- \mathcal{K}(\rho,\pi) +\log \epsilon>0 \right)  \leq \epsilon.
	\end{align*}
	Then taking the complementary and we obtain  with probability at least $1-\epsilon$ that:
	\begin{align*}
	\forall \rho, \quad (\lambda-\frac{C\lambda^2 }{2n(1-\lambda/n)}) \int  
	[R( \beta )-R^* ]\rho (d\beta ) 
	\leq 
	\lambda \int 
	[r_n( \beta )-r_n^* ]  \rho (d\beta ) + \mathcal{K}(\rho,\pi) + \log 
	\frac{1}{\epsilon}.
	\end{align*}
	Now, note that as $ r^h_n\geq r_n $,
	\begin{align*}
	\lambda \left[ \int r_n d\rho - r_n^* \right] + 
	\mathcal{K}(\rho,\pi)+\log\frac{1}{\epsilon}
\leq  
	\lambda \left[ \int r^h_nd\rho + \frac{1}{\lambda} 
	\mathcal{K}(\rho,\pi) 
	\right] - \lambda r_n^*  +\log\frac{1}{\epsilon} 
	.
	\end{align*}
	As it stands for all $\rho$ then the right hand side can be minimized and, from Lemma \ref{lemma_donvara}, the minimizer over $\mathcal{P}(\mathbb{R}^{d}) $ is $\hat{\rho}_\lambda$.	Thus we get, when $\lambda<2n/(C+2) $,
	\begin{align*}
	\int R d\hat{\rho}_\lambda 
	\leq 
	R^*  + \frac{1}{1-\frac{C\lambda }{2n(1-\lambda/n)}} 
	\left\lbrace 
	\inf_{\rho \in \mathcal{P}(\mathbb{R}^{d}) } 
	\left[ \int 
	r^h_nd\rho + \frac{1}{\lambda} \mathcal{K}(\rho,\pi) 
	\right] - r_n^*  + \frac{1}{\lambda} 
	\log\frac{1}{\epsilon} 
	\right\rbrace
	.
	\end{align*}

	\textit{Step 2:}	
	\\
From \eqref{eq_boundforhinge}, we have that,
	\begin{align*}
	\int r^h_n (\beta) \rho (d\beta) 
\leq
	r^h_n(\beta^{*}) 
	+ 	
	C_{\rm x}	\int 
	\|  \beta - \beta^{*}\|_2 
	\rho (d\beta)
	.
	\end{align*}
	And for $ \rho = p_0  $, as in \eqref{eq_boundforhinge}, and using Lemma \ref{lema_boundfor_ell2},
	\begin{align*}
	\int 
	\|  \beta - \beta^{*}\|_2 
	p_0 (d\beta)
	\leq 
	2 \tau \sqrt{d}
	. 
	\end{align*}
	From Lemma \ref{lema_boundof_KL}, we have that
	$$
	\mathcal{K}( p_0 ,  \pi) 
	\leq 
	4\|\beta^*\|_0  \log \left( 
	\frac{C_1 }{\tau \|\beta^*\|_0}
	\right)
	+
	\log(2)
	.
	$$
From Assumption \ref{assume_bound_on_thetruebayes}, as   $r^h_n(\beta^*) \leq (1+C')r^*_n $, we have that
	\begin{align*}
	\int R d\hat{\rho}_\lambda 
	\leq 
	R^* + 
	\frac{1}{1-\frac{C\lambda }{2n(1-\lambda/n)}} 
	\left\lbrace 
	C' r_n^*
	+
	C_{\rm x} 2 \tau \sqrt{d}
	+ 
	\frac{	4\|\beta^*\|_0  \log \left( 
		\frac{C_1 }{\tau \|\beta^*\|_0}
		\right)
		+	\log(2) }{\lambda} 
	+
	\frac{1}{\lambda} \log\left(\frac{1}{\epsilon}\right) 
		\right\rbrace
	.
	\end{align*}
Taking $\lambda = 2 n/(3C + 2) $,
	we obtain:
	\begin{align*}
&	\int R d\hat{\rho}_\lambda  
	\leq 	
	R^*
	+ 
	\\
&	\frac{3}{2} 
\left\lbrace 
C' 	r_n^*
+
C_{\rm x} 2 \tau \sqrt{d}
+ 
\frac{ (3C + 2) \left[ 4\|\beta^*\|_0  \log \left( 
	\frac{C_1 }{\tau \|\beta^*\|_0}
	\right)
	+	\log(2)\right] }{2n} 
+
\frac{(3C + 2) \log\left(1/\epsilon\right)}{2n} 
\right\rbrace
	.
	\end{align*}
	Then, we use Lemma~\ref{lem_theo1}, with probability at least $ 1-2\epsilon $, to obtain that
	\begin{align}
&	\int R d\hat{\rho}_\lambda 
	\leq 
	(1+3C' ) R^*  
	+ \nonumber
	\\
&	\frac{3}{2} 
\left\lbrace 	
C' \frac{1}{n\varsigma }\log \frac{1}{\epsilon}
+
C_{\rm x} 2 \tau \sqrt{d}
+ 
\frac{ (3C + 2) \left[ 4\|\beta^*\|_0  \log \left( 
	\frac{C_1 }{\tau \|\beta^*\|_0}
	\right)
	+	\log(2)\right] }{2n} 
+
\frac{(3C + 2) \log\left(1/\epsilon\right)}{2n} 
\right\rbrace
\label{eq_used_forknwon_sstatr_fastrate}
	.
	\end{align}
	By taking $ \tau = 1/(n\sqrt{d}) $, we can obtain that
	\begin{align*}
	\int R d\hat{\rho}_\lambda 
	\leq 
	(1+3C' ) R^*  
	+
\mathcal{C}_{C,C',C_{\rm x}} \frac{ 
	\|\beta^*\|_0  \log \left( 
	\frac{n\sqrt{d} C_1 }{ \|\beta^*\|_0}
	\right)
 + \log\left(1/\epsilon\right)
 }{n} 
	,
	\end{align*}
	where $ \mathcal{C}_{C,C',C_{\rm x}}  $ is a universal constant depending only on $ C,C',C_1, C_{\rm x} $.
	The proof is completed.
	
\end{proof}

\begin{proof}[\textbf{Proof of Theorem \ref{thrm_contraction}}]
	Let $\varepsilon_n $  be any sequence in $(0,1) $ such that $\varepsilon_n\rightarrow 0 $ when $n\rightarrow\infty$.	From \eqref{eq_to_contraction}, we have that
	\begin{align*}
	\mathbb{E} \Biggl[  \int \exp \left\{ 
	(\lambda- \frac{C\lambda^2 }{2n(1-\lambda/n)} )[R( \beta )-R^* ] 
	-\lambda[r_n( \beta )-r_n^* ]
	- 
	\log \left[\frac{d\hat{\rho}_{\lambda}}{d \pi} (\beta)  \right]
	- \log\frac{1}{\varepsilon_n}
	\right\}
	\hat{\rho}_{\lambda}(d \beta)
	\Biggr]
	\leq 
	\varepsilon_n
	.
	\end{align*}
	Using Chernoff's trick, i.e. using $\exp(x) \geq 1_{\mathbb{R}_{+}}(x)$, this gives:
	$$
	\mathbb{E} \Bigl[ 
	\mathbb{P}_{\beta \sim \hat{\rho}_{\lambda}} 
	(\beta \in \Omega_n ) \Bigr]
	\geq 1- \varepsilon_n
	$$
	where
	$$
	\Omega_n
	= 
	\left\{\beta : 	
	(\lambda- \frac{C\lambda^2 }{2n(1-\lambda/n)} )
	[R( \beta )-R^* ] 
	-\lambda[r_n( \beta )-r_n^* ]   
	\leq      
	\log \left[\frac{d\hat{\rho}_{\lambda}}{d \pi} (\beta)  \right]
	+ \log\frac{2}{\varepsilon_n} \right\}.
	$$
	Using the definition of $\hat{\rho}_\lambda $ and noting that $ r_n \leq r^h_n $, for $ \beta \in \Omega_n $ we have
	\begin{align*}
	\Bigl(
	\lambda- \frac{C\lambda^2 }{2n(1-\lambda/n)} 
	\Bigr)
	\Bigl[ R(\beta) - R^* \Bigr]
	&\leq  
	\lambda\Bigl( r_n(\beta) - r_n^* \Bigr)  +       \log \left[\frac{d\hat{\rho}_{\lambda}}{d \pi} (\beta)  \right]
	+ \log\frac{2}{\varepsilon_n}
	\\
	&    \leq  
	\lambda\Bigl( r^h_n(\beta) - r_n^* \Bigr)  +       \log \left[\frac{d\hat{\rho}_{\lambda}}{d \pi} (\beta)  \right]
	+ \log\frac{2}{\varepsilon_n}
	\\
	& \leq 
	- \log\int\exp\left[
	-\lambda r^h_n (\beta) \right]\pi({\rm d} \beta) - \lambda r_n^*
	+ \log\frac{2}{\varepsilon_n}
	\\
	& = \lambda\Bigl( \int r^h_n (\beta) \hat{\rho}_{\lambda}({\rm d} \beta) - r_n^* \Bigr)  +    \mathcal{K}(\hat{\rho}_\lambda,\pi)
	+ \log\frac{2}{\varepsilon_n}
	\\
	& = 
	\inf_{\rho} \left\{ 
	\lambda\Bigl( \int r^h_n (\beta) \rho({\rm d}\beta) - r_n^* \Bigr)  +    \mathcal{K}(\rho,\pi)
	+ \log\frac{2}{\varepsilon_n} \right\}.
	\end{align*}
	We upper-bound the right-hand side exactly as Step 2 in the proof of Theorem~\ref{thm_fastrate} (with Lemma \ref{lem_theo1}). The result of the theorem is followed.
	
\end{proof}

\subsection{Proof of auxiliary lemmas}

\begin{lemma}
	\label{lem_theo1}
	For $\epsilon \in (0,1) $, with probability at least $1-\epsilon$, we have, for every $ \varsigma \in (0,1) $, that
	\begin{align*}
	r_n^* 
	\leq 
(1+\varsigma)	R^* +\frac{1}{n\varsigma }\log \frac{1}{\epsilon}
	.
	\end{align*}
	or we can have $ 	r_n^* 
	\leq 
2R^* +\frac{1}{n\varsigma }\log \frac{1}{\epsilon} $.
\end{lemma}

\begin{proof}
	Let $ \varsigma \in (0,1)$, we have that
	\begin{align*}
	\mathbb{E}\left(\exp [ \varsigma n r_n^* ]\right) 
&	= 
	\prod_{i=1}^n  
	\mathbb{E}\left(
	\exp \left[ \varsigma \mathbbm{1}_{(Y_{i}  ( \beta^{*\top}  x_{i} )<0)} \right] \right) 
	\\
	&	= 
	\prod_{i=1}^n  
	\mathbb{E}\left\{
	\exp \left[ \varsigma \mathbbm{1}_{(Y_{i}  ( \beta^{*\top}  x_{i} )<0)} 
	+
	0(1 - \mathbbm{1}_{(Y_{i}  ( \beta^{*\top}  x_{i} )<0)} )
	\right]
	\right\} 
	\\
&	\leq 
	\prod_{i=1}^n  
	\left\{
	e^\varsigma \mathbb{E}\left[ 
	\mathbbm{1}_{(Y_{i} ( \beta^{*\top}  x_{i} )<0)} \right] 
	+ \left(
	1- \mathbb{E}\left[ 
	\mathbbm{1}_{(Y_{i} ( \beta^{*\top}  x_{i} )<0)} \right]
	\right)
	\right\} 
	\\
&	\leq 
	\prod_{i=1}^n  
	\left( e^\varsigma R^* +1 - R^*
	\right) 
=	\prod_{i=1}^n  
\left(  R^*(e^\varsigma -1) +1
\right) 
	\\
&	\leq 
\prod_{i=1}^n  
\exp \left(  R^*(e^\varsigma -1)
\right) 
= \exp \left( n R^*(e^\varsigma -1)
\right) 
	.
	\end{align*}
	Thus we obtain, for $\epsilon \in (0,1)$:
	\begin{align*}
	\mathbb{E}\left[ \exp \left( \varsigma nr_n^*  -  n R^* (e^\varsigma -1)  -\log 
	\frac{1}{\epsilon} \right)\right] \leq \epsilon.
	\end{align*}
	Now, using Markov's inequality that $ \mathbb{P}(W>0) \leq \mathbb{E}[e^W] $ for any $ W $, we get that 
	\begin{align*}
	\varsigma nr_n^*  -  (e^\varsigma -1)  n R^*  -\log 
	\frac{1}{\epsilon}
	\leq 0 ,
	\end{align*}
	with probability at least $ 1-\epsilon $. Thus, the result of the lemma is obtained by noting that $ e^\varsigma \leq 1 + \varsigma +\varsigma^2, \varsigma \in (0,1) $.
	
\end{proof}

%%% the translate prior
\begin{definition}
We define the following distribution as a translation of the prior $ \pi $,
\begin{equation}
\label{eq_specific_distribution}
p_0(\beta) 
\propto 
\pi (\beta - \beta^*)\mathbbm{1}_{B_1(2d\tau)} (\beta - \beta^*).
\end{equation}
\end{definition}
It is worth highlighting that given $ \| \beta^* \|_1 \leq C_1 - 2d\tau $, when the condition $ \beta - \beta^*\in B_1(2d\tau) $ holds, it implies that $ \beta \in B_1(C_1) $. Consequently, the distribution $ p_0 $ is absolutely continuous with respect to the prior distribution $ \pi $.

\begin{lemma}
	\label{lema_boundfor_ell2}
	Let $p_0 $ be the probability measure defined by (\ref{eq_specific_distribution}). If
	$d\geq 2$ then
	$$
	\int_\Lambda \| \beta- \beta^* \|^2 p_0(d \beta)
	\leq
	 4d\tau^2 
	 .
	$$
\end{lemma}
\begin{proof}
	First, we have that
	$$
	\int_\Lambda \| \beta- \beta^* \|^2 p_0(d \beta)
	=
	d\int_\Lambda ( \beta_1- \beta_1^*)^2p_0(d \beta)
	.
	$$
Using Lemma 2 from \cite{dalalyan2012mirror} we get
	\begin{eqnarray*}
	\int_\Lambda ( \beta_1- \beta_1^*)^2p_0(d \beta)
	\le 
	4\tau^2
	\end{eqnarray*}
	and the desired inequality follows.
	
\end{proof}

\begin{lemma}
	\label{lema_boundof_KL}
	Let $p_0$ be the probability measure defined by (\ref{eq_specific_distribution}). Then
	$$
	KL(p_0,\pi)
	\leq
	4 s^* \log \left(\frac{C_1 }{\tau s^*}\right)
	+
	\log(2)
	.
	$$
\end{lemma}
\begin{proof}
From Lemma 3 in \cite{dalalyan2012mirror}, we have that
	$$
KL(p_0,\pi)
\leq
4\sum_{j=1}^d \log(1+|\beta_j^*|/\tau)
+
\log(2)
.
$$
Then, from Corollary 1 in \cite{dalalyan2012mirror} (that using Jensen's inequality), we have that
	$$
\frac{1}{s^*}\sum_{j=1}^d \log(1+|\beta_j^*|/\tau)
\leq
 \log \left( 
1+ \frac{\|\beta^*\|_1}{\tau s^*}
 \right)
.
$$
As $ \| \beta^* \|_1 \leq C_1 - 2d\tau $, we then have that
	$$
 \log \left( 
1+ \frac{\|\beta^*\|_1}{\tau s^*}
\right)
\leq
 \log \left( 
1+ \frac{C_1 - 2d\tau}{\tau s^*}
\right)
\leq
\log \left( 
\frac{C_1 }{\tau s^*} -1
\right)
\leq
\log \left( 
\frac{C_1 }{\tau s^*}
\right)
,
$$
by noting that $ \|\beta^*\|_0 \leq s^*\leq n < d $.
Thus, we obtain the desired  result.

\end{proof}

\subsection{Proofs for Section \ref{sc_known_sstar}}
\label{sc_proof_knownstars}
\begin{proof}[\bf Proof of Proposition~\ref{thm_slowrate_known_sstars}]		
From the proof of Theorem \ref{thm_slowrate}, inequality \eqref{eq_used_forknwon_sstatr}, with probability at least $ 1-2\epsilon $, we have that
	\begin{align*}
	\int R d\hat{\rho}_\lambda 
	\leq 
	(1+2C')R^*  
	+
C'	\frac{1}{n\varsigma }\log \frac{1}{\epsilon}
	+
	C_{\rm x} 2 \tau \sqrt{d}
	+ 
	\frac{	4s^* \log \left(\frac{C_1 }{\tau s^*}\right)
		+
		\log(2) }{\lambda} 
	+
	\frac{\lambda }{2n} 
	+
	\frac{1}{\lambda} \log\left(\frac{1}{\epsilon}\right) 
	.
	\end{align*}
	By taking now  $ \tau = s^*/( n\sqrt{d} ) $ and $ \lambda = \sqrt{n s^* \log (de/s^*) } $, we obtain that 
	\begin{align*}
&	\int R d\hat{\rho}_\lambda 
	\leq 
	\\
&	(1+2C')R^*  
	+
	C_{\rm x} \frac{2s^*}{n}
	+  	
	\frac{	4s^* \log \left(\frac{n\sqrt{d} C_1 }{ s^* s^*}\right)
	}{\sqrt{n s^* \log (de/s^*) } } 
	+
	\frac{\sqrt{ s^*\log (de/s^*) }}{2\sqrt{n}} 
	+
	\left( \frac{1}{\sqrt{n s^*\log (de/s^*) }} + \frac{C'}{n\varsigma } \right) \log\left(\epsilon^{-1} \right) 
	.
	\end{align*}
By noting that 
\begin{align}
\label{eq_noticefor_sharprate}
\frac{n\sqrt{d} }{ s^* s^*}
=
\frac{n }{ s^*e \sqrt{d}}
\frac{de }{  s^*}
\leq
\left(\frac{de }{  s^*}\right)^2
\end{align}	
	Therefore, we can obtain that
	\begin{align*}
	\int R d\hat{\rho}_\lambda 
	\leq 
(1+2C')R^*  
	+
c \frac{\sqrt{s^* \log \left(de/s^*\right) }  }{\sqrt{n}} 
	+ 
c	\frac{\log\left(1/\epsilon\right)}{\sqrt{n s^* \log (de/s^* ) }} 
	,
	\end{align*}
	where $ c $ is a universal constant depending only on $ C_1,C', C_{\rm x} $.
	The proof is completed.
	
\end{proof}

\begin{proof}[\bf Proof of Proposition~\ref{thm_fastrate_known_sstars}]
From the proof of Theorem \ref{thm_fastrate}, inequality \eqref{eq_used_forknwon_sstatr_fastrate}, with probability at least $ 1-2\epsilon $, we have that	\begin{align*}
	&	\int R d\hat{\rho}_\lambda 
	\leq 
(1+3C') R^*  
	+
	\\
	&	\frac{3}{2} 
	\left\lbrace 		\frac{C'}{n\varsigma }\log \frac{1}{\epsilon}
	+
	C_{\rm x} 2 \tau \sqrt{d}
	+ 
	\frac{ (3C + 2) \left[ 4s^*  \log \left( 
		\frac{C_1 }{\tau s^*}
		\right)
		+	\log(2)\right] }{2n} 
	+
	\frac{(3C + 2) \log\left(1/\epsilon\right)}{2n} 
	\right\rbrace
	.
	\end{align*}
	By taking now $ \tau = s^*/(n\sqrt{d}) $, we obtain that
	\begin{align*}
	\int R d\hat{\rho}_\lambda 
	\leq 
(1+3C') R^*  
	+
	\mathcal{C}_{C,C',C_{\rm x}} \frac{ 
		s^*  \log \left( 
		\frac{n\sqrt{d} C_1 }{ s^* s^*}
		\right)
		+ \log\left(1/\epsilon\right)
	}{n} 
	,
	\end{align*}
	and using the notice in \eqref{eq_noticefor_sharprate}, we obtain 
	\begin{align*}
\int R d\hat{\rho}_\lambda 
\leq 
(1+3C') R^*  
+
\mathcal{C}_{C,C',C_{\rm x}} \frac{ 
	s^*  \log \left( 
	\frac{de  }{  s^*}
	\right)
	+ \log\left(1/\epsilon\right)
}{n} 
,
\end{align*}	
	where $ \mathcal{C}_{C,C',C_{\rm x}}  $ is a universal constant depending only on $ C,C_1,C', C_{\rm x} $.
	The proof is completed.
	
\end{proof}

\newpage
{\footnotesize 
%	\bibliographystyle{plain}
%	\bibliography{refs_sparselogistic}

}

\end{document}